# High field magneto-transport study of $YBa_2Cu_3O_7$:$Ag_x$ (x = 0.00–0.20)

Poonam Rani, Anand Pal and V.P.S Awana*

National Physical Laboratory (CSIR), Dr. K.S Krishnan Marg, New Delhi-110012, India

We report high field (up to 13 Tesla) magneto transport [$\rho(T)H$] of $YBa_2Cu_3O_7$ (YBCO):$Ag_x$ (x= 0.0, 0.1 and 0.2) composites. The transport properties are significantly improved by Ag doping on the insulating grain boundaries of YBCO. Pure and Ag diffused YBCO superconducting samples are synthesised through solid state reaction route. Both pure and Ag doped YBCO are superconducting at below 90K. Though, the $T_c$ ($\rho$=0) of YBCO:Ag samples under applied field of 13 Tesla is around 65K, the same is 45K for pure YBCO under same applied field. The upper critical field [$H_{c2}(0)$], being estimated from $\rho(T)H$ is around 70Tesla for pristine sample, and is above 190Tesla for Ag doped samples. The boarding of the resistive transition under applied magnetic field is comparatively less and nearly single step for Ag doped samples, while the same is clearly two step and relatively much larger for the pristine YBCO. The resistive broadening is explained on the basis of changed inter-granular coupling and thermally activated flux flow (TAFF). The TAFF activation energy ($U_0$) is found to be linear with applied magnetic field for all the samples, but with nearly an order of magnitude less value for the Ag doped samples. Summarily, it is shown that inclusion of Ag significantly improves the superconducting performance of YBCO:Ag composites, in particular under applied field.

*Corresponding author: e-mail – awana@mail.npindia.org
Tel.: +91 11 45609357; Fax; +91 11 45609310.

**Introduction**

Since the discovery of high $T_c$ superconductivity at above 90K [1] in YBa2Cu3O7 (YBCO), lot of efforts have been made to improve the superconductivity and granular coupling of bulk YBCO [2-3]. The researchers have tried to improve the granular coupling of YBCO by making composites of the superconductor with impurities like Ag, Au etc. [2-5]. Very recently, we reported that the Ag doping enhances the inter-granular couplings of YBCO through grain alignment, grain boundary doping and minimization of grain-boundary area [6]. This was demonstrated by detailed AC susceptibility and SEM (scanning electron microscopy) studies on YBCO:$Ag_x$ composites [6]. The controlled addition of Ag in between the grains of YBCO results in better interconnection, which improves weak links and leads to the enhancement in the superconducting properties.

In this short communication, we further extend our previous study [6], by reporting the high field (up to 13Tesla) magneto-transport of the previously studied $YBa_2Cu_3O_7$ (YBCO): $Ag_x$ (x= 0.0, 0.1 and 0.2) composites. We study the variation in critical temperature ($T_c$), upper critical field $H_{c2}$ and flux pinning behaviour of YBCO with concentration of Ag. Inclusion of Ag showed slight increase in $T_c$ and interestingly substantial enhancement in upper critical field $H_{c2}$. $H_{c2}$ (upper critical field) is an important parameter of a superconductor, which is useful for practical applications of these materials. The improvement in the value of $H_{c2}$ demonstrated the enhancement of superconducting grains coupling. Here we measured the resistivity transition under high magnetic field (up to 13Tesla). The breaking of resistive transition under magnetic field takes place in two steps i.e., (a) due to percolation nature between individual superconducting grains and (b) between two superconducting grains through the grain boundary junctions. Part (a) is seen as the onset of superconducting transition due to the superconductivity in individual grains and (b) is the long transition tail due to the couplings regimes between grains or connective nature of grains [7]. The temperature derivative of resistivity ($d\rho/dT$) provides information about fluctuation conductivity and the characteristics structure of two stage transitions [7]. The $d\rho/dT$ plot shows a high peak at the bulk $T_c$ and secondary hump is seen at lower temperature side. The secondary hump is characteristic of the connective nature of the grains pushing the resistance towards zero through grain boundary weak links [7-10]. Another, important information one can get from the magneto-transport studies in the superconducting transition regime is the Dew-Hughes phenomenon of thermal activation flux flow, (*TAFF*) [11]. Below $T_c$, the *HTSc* compounds show an ohmic resistance under magnetic field which is called the thermal active flux and is related to flux creep model of Anderson and Kim [12]. In this paper we report that high magnetic field transport measurements of Pure and Ag doped YBCO. It is found that the inter-granular coupling is improved by addition of Ag in YBCO. Limited addition of Ag enhances the upper critical field $H_{c2}$ by better inter connectivity of the superconducting grains. The $H_{c2}(0)$ is increased by three fold from around 70Tesla for pure YBCO to above 190Tesla for Ag added samples. Our results clearly opens up the opportunity for high field superconducting applications of YBCO:Ag composites.

## Experimental

The samples of $YBa_2Cu_3O_{7-\delta}+Ag_x$ series are prepared through conventional solid state reaction route with nominal composition x= 0, 0.10 and 0.20wt%. Details are given ref. 6. All the samples were characterized by the X-ray powder diffraction technique using Rigaku X-ray diffractometer (Cu-$K_\alpha$ line) and Rietveld analysis of all samples was performed using Fullprof program. Detailed resistivity measurements under applied magnetic field of up to 13 Tesla are done on Physical Properties Measurement System (Quantum Design-USA PPMS-14Tesla). The SEM images of the samples were taken using ZEISS EVO MA-10 Scanning Electron Microscope.

## Results and Discussion

The Rietveld fitted *XRD* refinement of $YBa_2Cu_3O_7+Ag_x$ (x= 0, 0.10, 0.20) samples showed that all the samples are crystallized in single phase having orthorhombic structure within *Pmmm* space group having lattice parameter $a$= 3.825(2) Å, $b$=3.889(8) Å and $c$= 11.684(4) Å. Although no impurity peaks were detected in the pure sample, some small impurity peaks of Ag were observed in Ag doped YBCO samples [6]. The lattice parameters and orthorhombicity remained nearly invariant for pure and Ag added samples.

Figure 1 shows the temperature dependence of resistivity of $YBa_2Cu_3O_7+Ag_x$ (x= 0, 0.10, 0.20) samples. The data is normalized to unity at 300K. Above critical temperature linear resistivity curve is observed for all samples, exhibiting their metallic behavior. Relatively high normal resistivity is observed in pure YBCO than the doped samples. Also the slope of metallic resistivity is more for Ag doped YBCO than pure YBCO. For example, the normalized resistivity before superconducting onset (~ 91K) is though decreased to around 55% for pure YBCO, the same is below 35% for Ag added samples. This shows that Ag added samples are more metallic than pure YBCO, seemingly due to better transport channels. Below the superconducting onset temperature of around 91K, all samples exhibit sharp superconducting transition with their $T_c$ ($\rho$ = 0) of around 88-89K.

Figures 2 (a-c) show the temperature dependence of normalized resistivity $\rho/\rho_{95}$ of studied $YBa_2Cu_3O_7+Ag_x$ (x= 0, 0.10, 0.20) samples under magnetic field up to 13Tesla. The field is applied perpendicular to the current flow direction. The magneto-transport (up to 13Tesla) of pure YBCO sample is shown in Fig. 2(a). At zero field, the critical temperature at zero resistivity $T_c$ ($\rho$= 0) for the pure YBCO sample is around 88K. With applied field though the superconducting onset of around 91K remains unaltered, the $T_c$ ($\rho$= 0) is decreased to lower temperatures. Under applied field of 13Tesla, the $T_c$ ($\rho$= 0) of pure YBCO is decreased to below 45K. The decrease of $T_c$ ($\rho$= 0) of pure YBCO is accompanied with a two step transition. In zero applied field, the superconducting transition is single step with $T_c$ ($\rho$= 0) of 88K for pure YBCO. The broadening of transition with clear two step nature in small fields of 0.1Tesla is reminiscent of insulating grain boundaries between superconducting grains of pure YBCO [4-6, 7]. One needs to overcome the insulating grain boundaries for high field practical applications of YBCO superconductor [4, 7]. This question is haunting the scientific community from over two decades since the invention of high $T_c$ superconductivity of cuprates. Figures (2b and 2c) show the temperature dependence of normalized resistivity $\rho/\rho_{95}$ of studied YBCO:Ag10 and YBCO:Ag20 samples under applied magnetic fields of up to 13Tesla. Though the onset of superconducting transition remains close to above 90K, the $T_c$ ($\rho$= 0) is shifted to low temperature. The $T_c$ ($\rho$= 0) of YBCO:Ag10 and YBCO:Ag20 samples under applied field of 13Tesla is around 65K and 67K respectively. Interestingly for pure YBCO the $T_c$ ($\rho$= 0) is 45K under same applied field of 13Tesla. This shows that addition of Ag has significantly improved the superconducting performance of YBCO superconductor. Further more striking is the fact that though the superconducting transition of pure YBCO breaks clearly in two steps under applied field (Fig. 2a), the same is nearly single step for both Ag10 and Ag20 samples. As mentioned earlier, the broadening of superconducting transition with clear two step nature under applied is reminiscent of insulating grain boundaries between superconducting grains of a superconductor. It is clear that the granular coupling is significantly enhanced for Ag added YBCO samples. This corroborates our previous study related to detailed AC susceptibility, indicating improved granular coupling for YBCO:Ag composites [6].

The breaking of superconducting transition under magnetic field can be better viewed by plotting the derivative resistivity with temperature (d$\rho$/d$T$) in superconducting transition regime. The d$\rho$/d$T$ plots for YBCO-Pure, YBCO:Ag10 and YBCO:Ag20 samples are depicted in Figs 3(a), (b) and (c) respectively. The first peak ($T_{P1}$) in d$\rho$/d$T$ does appear due to the intra grain and the second one ($T_{P2}$) due to inter grain superconducting transitions. In zero field the transition is seen in nearly a single step due to the resistivity percolation through grain boundaries and only $T_{P1}$ is seen in d$\rho$/d$T$. However, once the field is applied the transition breaks in two parts and both $T_{P1}$ and $T_{P2}$ could be seen, this is the case for pure YBCO (Fig. 3a). Further, under applied field though the position (temperature) of $T_{P1}$ remains nearly invariant due to very high upper critical field of YBCO, the $T_{P2}$ shifts to lower temperature. In fact $T_{P2}$ (the inter grain coupling transition), is principally an *SIS* (superconducting – insulating – superconducting) junction in case of YBCO, which shifts to lower temperature with field due to decreased proximity effects or weakened Josephson tunnelling [2,7]. Interestingly for both YBCO:Ag10 and Ag20 samples the second peak ($T_{P2}$) is not visible even under applied fields of as high as up to 13Tesla, see figures 3(a) and (b). It seems as if the YBCO:Ag composite sample are like single domain and the inter grain transition is not visible [13,14]. It seems that the granular peak is disappeared on addition of Ag in YBCO. This clearly demonstrates the improvement in granular coupling of YBCO:Ag composites. This is in accordance with our recently published AC susceptibility results on same samples [6].

Figure 4 shows temperature dependency of resistive upper critical field using the data points where the resistivity is half of its normal state value. It is observed that all the samples show concave curvature (upward curvature) near the $T_c$ and afterward it shows linearity in the curve. The upward curvature increases with Ag doping. It increases with Ag doping but for 20% Ag doped sample a slight decrease is found. Werthamer, Helfand, and Hohenberg (WHH) theory gave a solution for linearized Gor'kov equations for $H_{c2}$ for bulk weakly coupled type-II superconductor, including effects of Pauli spin paramagnetism and spin–orbit scattering. Here we use simplified WHH equation to estimate $H_{c2}(T)$ without spin paramagnetism and spin orbit interaction which is given by

$$ln\frac{1}{t} = \Psi\left(\frac{1}{2} + \frac{\hbar}{2t}\right) - \Psi\frac{1}{2}$$

Where, t = $T/T_c$, W is the digamma function and ħ is given by,

$$\hbar = \frac{4\,H_{c2}}{\pi^2 T_c(-dH_{c2}/dT)_{T=T_c}}$$

Using slope of linear portion of the curve of experimental data and corresponding extrapolated $T_c$ values to $H_{c2}$ = 0 region, the $H_{c2}(T)$ is fitted up to 0K using simplified WHH model [7-9] and is given in Figure 4. The calculated $H_{c2}(0)$ is around 70Tesla for pristine sample, 193 Tesla for 10% Ag 191Teslafor 20% Ag doped YBCO samples. It is clear that both $(dH_{c2}/dT)_{Tc}$ and $H_{c2}(0)$ improves significantly with doping of Ag in YBCO system possibly due to better grains coupling. The resistive transitions for pure and Ag added YBCO show a broadening near $T_c$ (ρ=0) under magnetic field (shown in fig. 2) due to creeps of vortices [15,16], which are thermally activated and explained in terms of thermally activated flux flow (TAFF). The temperature dependence resistivity is given by Arrhenius equation [15],

$$\rho(T, B) = \rho_0 \exp(-U_0/k_B T)$$

where $\rho_0$ is the field independent parameter (here normal state resistance at 95 K ($\rho_{95}$) is taken as $\rho_0$), $k_B$ is Boltzmann's constant and $U_0$ is the TAFF activation energy, which can be obtained from the slope of the linear part of the ln $(\rho/\rho_0)$ versus $T^{-1}$plot. Figures 5(a), 5 (b) and 5 (c) show the ln $(\rho/\rho_0)$ versus $T^{-1}$ plots for the pure and Ag doped YBCO samples under applied magnetic field. The line with blue symbol shows the experimental data of ln$(\rho/\rho_0)$ versus $T^{-1}$ and the red line shows best fit to the same. The activation energy is obtained from the slope of the linear curves, which are plotted under different magnetic fields. As calculated from Fig. 5(a) the activation energy ($U_0/k_B$) value for pure YBCO ranges from 1975.04K to 533.73K in the magnetic of 0 to 13Tesla. For 10% and 20% Ag doped YBCO, the $U_0$ varies from 4793.21K to 1060.45K and 2165.84K to 772.45K respectively under magnetic field ranging from 0 to 13 Tesla. Clearly the activation energy is more for Ag doped YBCO samples. Figure 6 (a) and (b) depicts the SEM images of the pure and 10% Ag doped YBCO samples. It is clearly seen that the grains of pure sample are smaller than the Ag doped sample i.e. the grain size increases with Ag Doping. Also in case of 10% Ag doped sample the surface texture is improved with decreased porosity level in comparison of the pristine sample. Chunks of Ag are also seen on the grain boundaries of Ag doped YBCO sample. These observations indicate improved coupling of superconducting grains in case of Ag doped YBCO samples.

In conclusion our detailed $\rho(T)H$ studies on pure and Ag doped YBCO demonstrated that Ag inclusion improve significantly the superconducting properties of YBCO superconductor. Particularly the upper critical filed [$H_{c2}(0)$] is increased nearly three fold from around 70 Tesla to above 190 Tesla for Ag doped samples. Present results opens up the opportunity for high field superconducting applications of YBCO:Ag composites.

Authors acknowledge keen interest and encouragement of their Director Prof. R.C. Budhani. Poonam Rani is supported by CSIR research intern scheme. Anand Pal is financially supported from DAE-SRC outstanding researcher scheme. The work is supported from DAE-SRC outstanding researcher Award scheme to work on search for new superconductors.

**Figure Captions**

Figure 1: Normalised resistivity ($\rho/\rho_{300}$) plots for pure YBCO and YBCO:Ag composites.

Figure 2: Normalized resistivity ($\rho/\rho_{95}$) plots under applied magnetic field of up to 13 Tesla for (a) YBCO, (b) YBCO-Ag10, and (c) YBCO-Ag20

Figure 3: Derivative of resistivity with temperature ($d\rho/dT$) in superconducting transition regime for (a) YBCO, (b) YBCO-Ag10, and (c) YBCO-Ag20

Figure 4: Fittings of resistive upper critical field [$H_{C2}(T)$] using simplified WHH theory (using the data points where the resistivity is half of its normal state value) of $YBa_2Cu_3O_7$:Ag, $x$ = 0.00, 0.10 and 0.20 samples.

Figure 5: Fitted and observed Arrhenius plots in TAFF region for (a) YBCO, (b) YBCO-Ag10, and (c) YBCO-Ag20.

Figure 6: SEM micrographs of (a) pure YBCO and (b) 10% Ag doped YBCO.

**References**

1. M. K. Wu, J. R. Ashburn, C. J. Torng, P. H. Hor, R. L. Meng, L. Gao, Z. J. Huang, Y. Q. Wang, and C. W. Chu, Phy. Rev. Lett. **58**, 908 (1987)
2. J. Jung, I. Issac and M. Mohamed, Physical Review B **48**, 7526 (1993)
3. A.R. Jurelo, I. A. Castillo, J. Roa-Rojas , L.M. Ferreira, L. Ghivelder, P. Pureur, P. Rodrigues Jr., Physica C **311**, 133 (1999).
4. C. Lambert, J.M. Debierre, G. Albinet and J. Marfaing, Philosophical Magazine B **79**, 1029 (1999).
5. M. Tape, I. Avci, H. Kocoglu and D. Abukay, Solid State Communications **131**, 319 (2004).
6. P. Rani. R. Jha. V.P.S. Awana, J. Sup. & Novel Mag. **26**, L2347 (2013).
7. B. W. Veal, W. K. Kwok, A. Umezawa, G. W. Crabtree, J. D. Jorgensen, J. W. Downey, L. J. Nowicki, A. W. Mitchell, A. P. Paulikas and C. H. Sowers, Appl. Phys. Lett. **51**, 279 (1987).
8. N. P. Liyanawaduge, S. K. Singh, A. Kumar, R. Jha, B. S. B Karunarathne and V.P. S. Awana, Supercond. Sci. Technol. **25**, 035017 (2012).
9. N.P. Liyanawaduge, A. Kumar, R. Jha b, B.S.B. Karunarathne, and V.P.S. Awana, Journal of Alloys and Compounds **543**, 135 (2012).
10. J. Kumar, D. Sharma, P.K. Ahluwalia, V.P.S. Awana, Materials Chemistry and Physics **139**, 681 (2013).
11. Dew Hughes, Cryogenics **28**, 674 (1988).
12. P. W. Anderson and Y. B. Kim, Rev. Mod. Phys. **36**, 39 (1964).
13. Judith L. MacManus-Driscoll, Advanced Materials **9**, 457 (1997).
14. M. Murakami, M. Morita, K. Doi, K. Miyamoto, H. Hamada, Jpn. J. Appl. Phys. **28**, L1065 (1989).
15. A. Gurevich, Rep. Prog. Phys. **74**, 124501 (2011).
16. J. Jaroszynski J, F. Hunte, L. Balicas, Jo Youn-jung , I. Rai Cevic, A. Gurevich, D. C. Larbalestier, F.F. Balakirev, L. Fang, P. Cheng, Y. Jia, and H. H. Wen, Phys. Rev. B **78**, 174523 (2008).

Figure 1

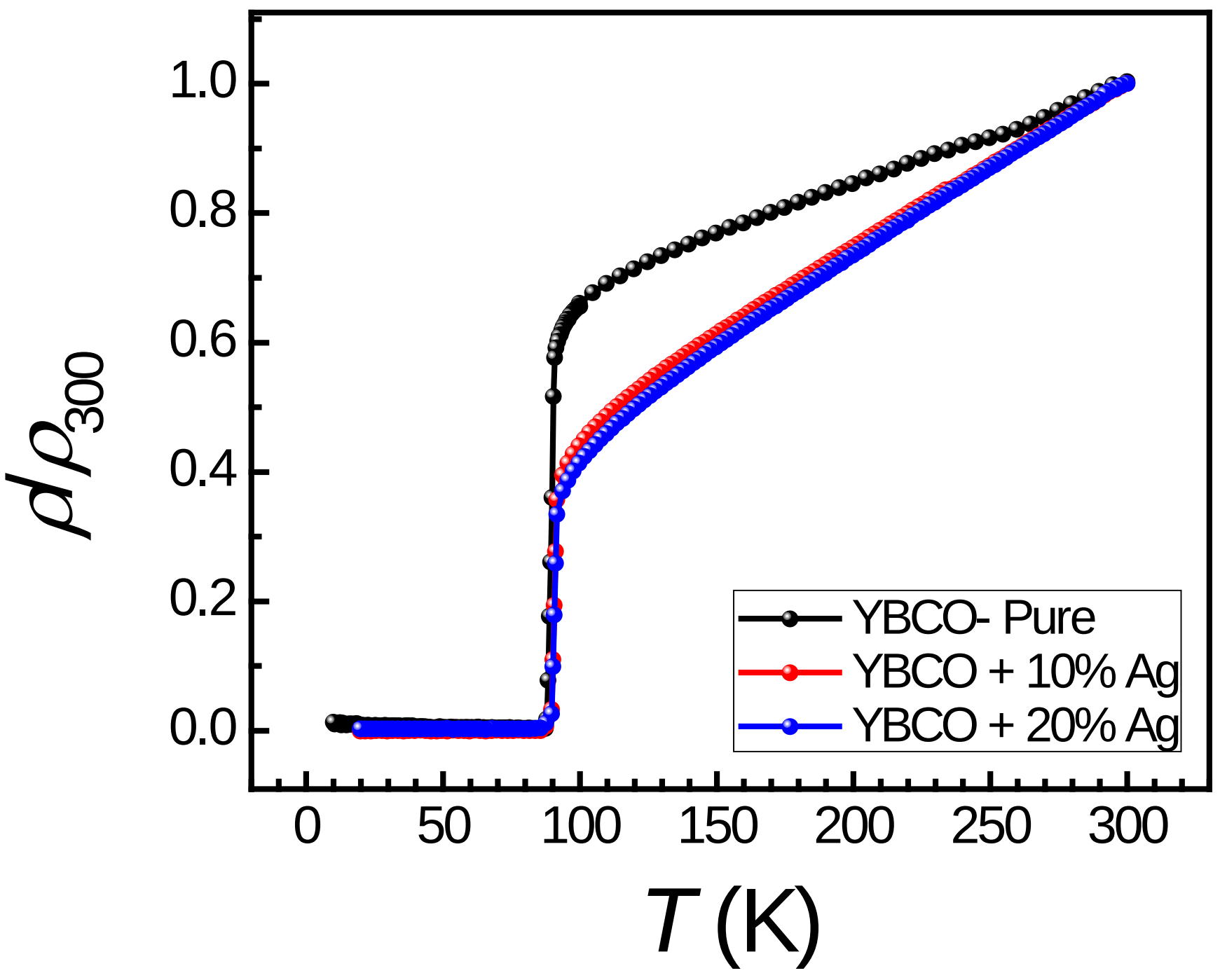

Figure 2 (a)

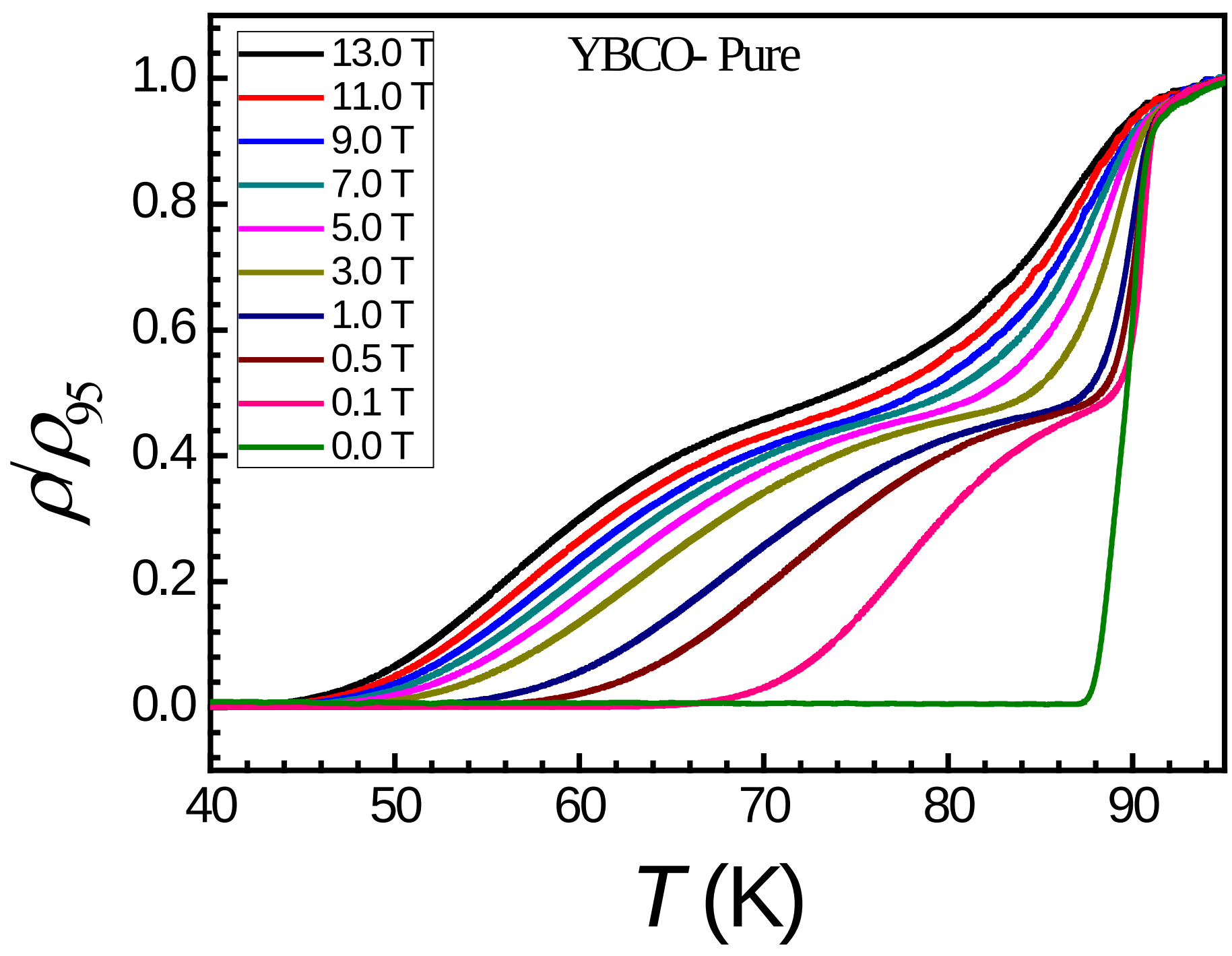

Figure 2(b)

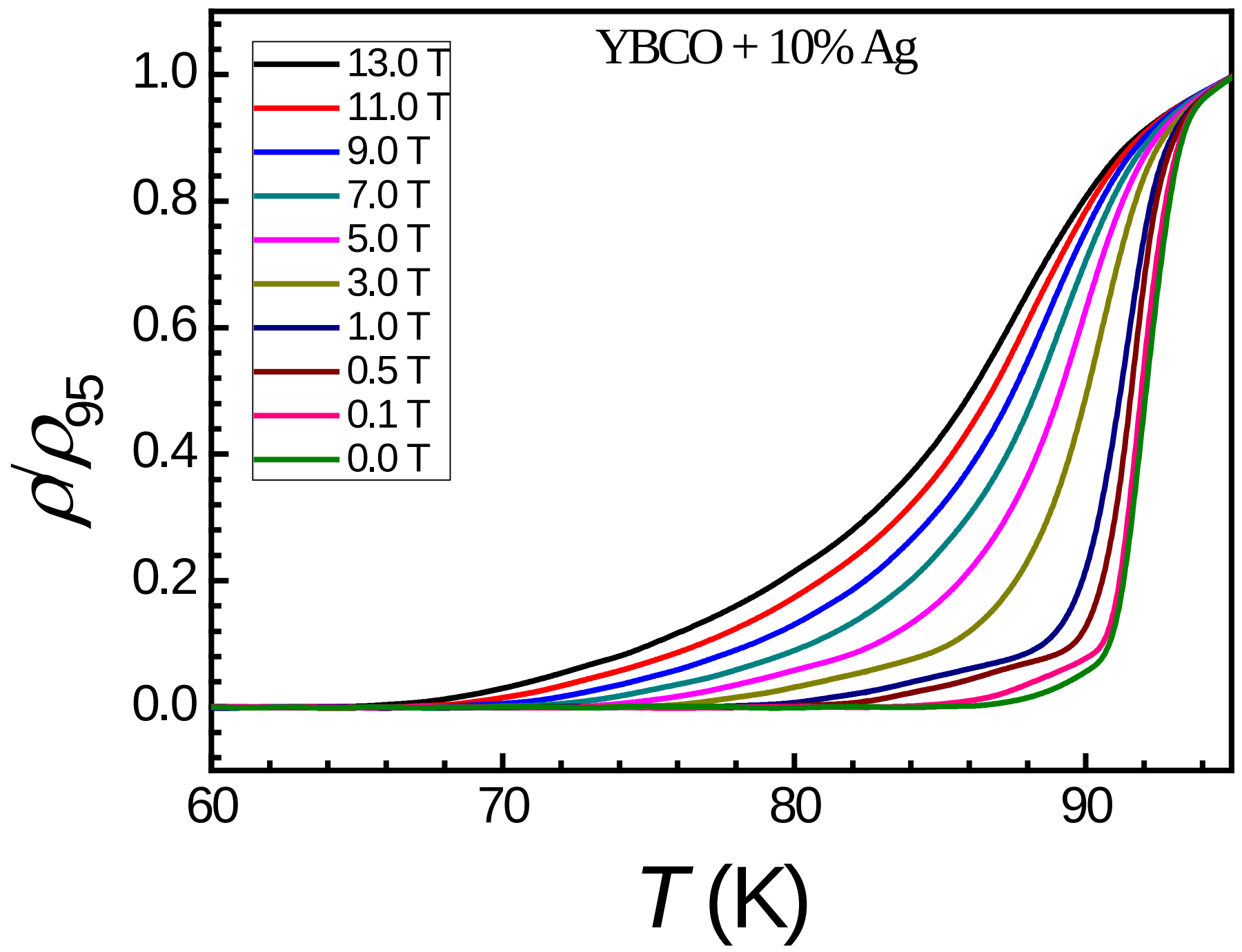

Figure 2 (c)

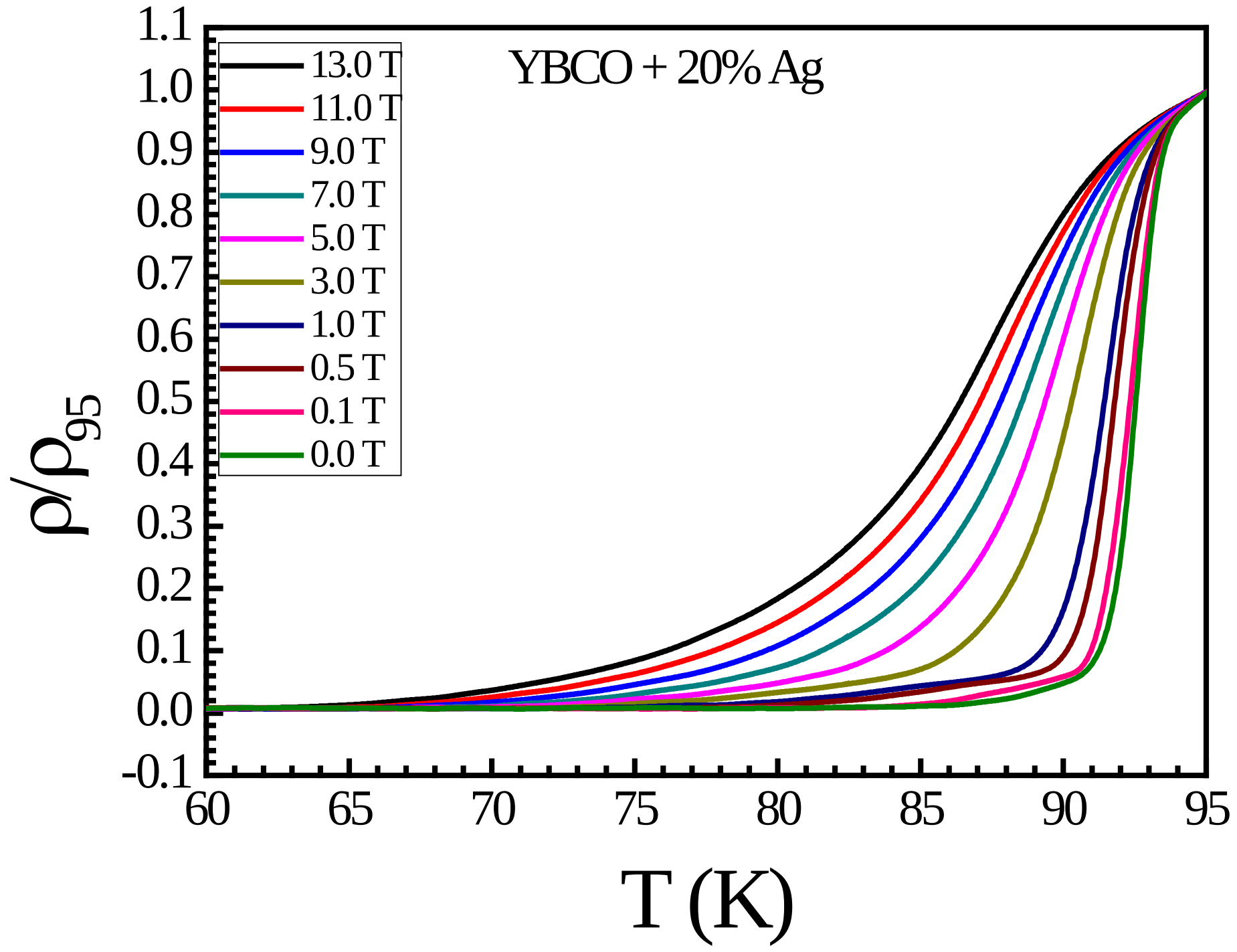

Figure 3 (a)

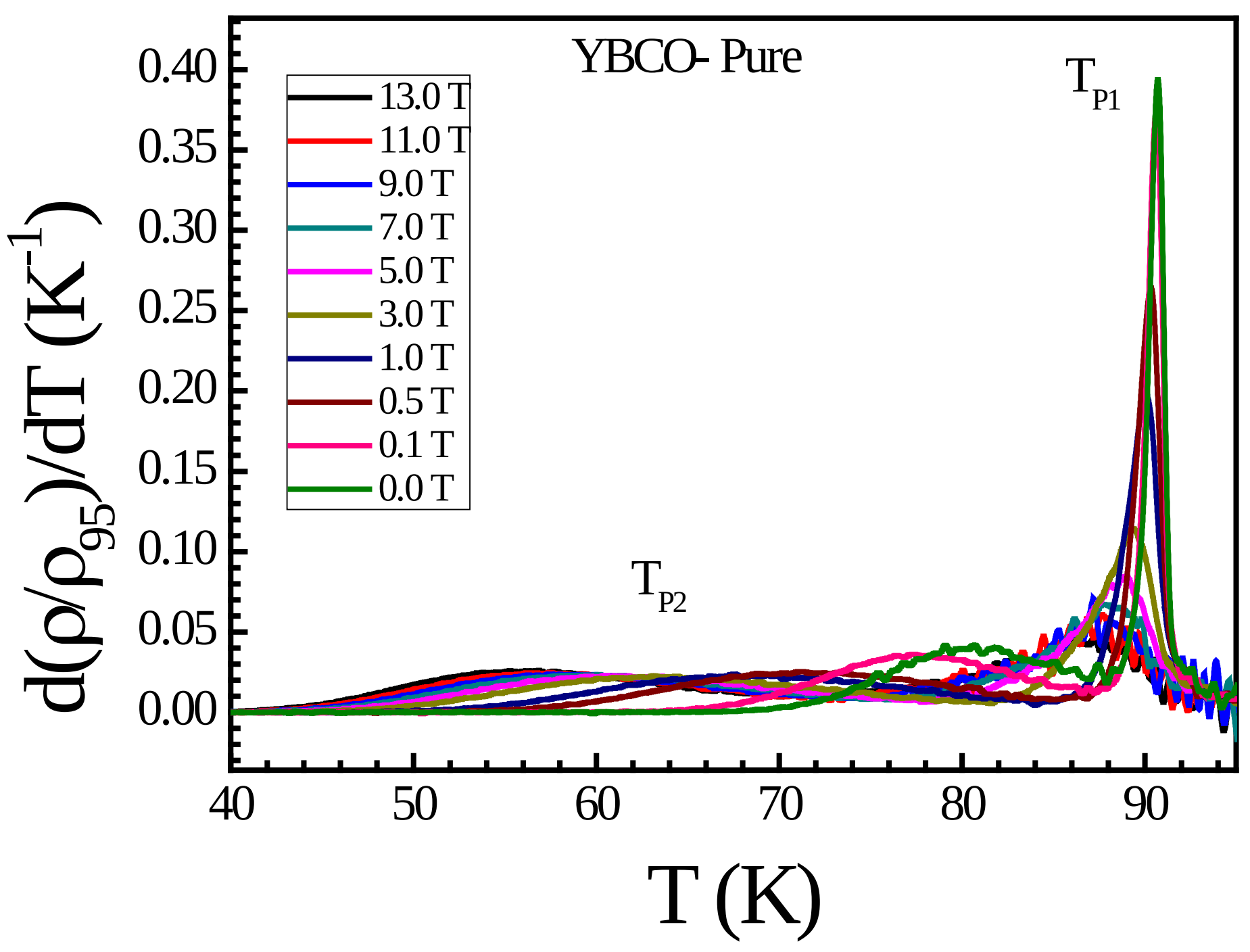

Figure 3 (b)

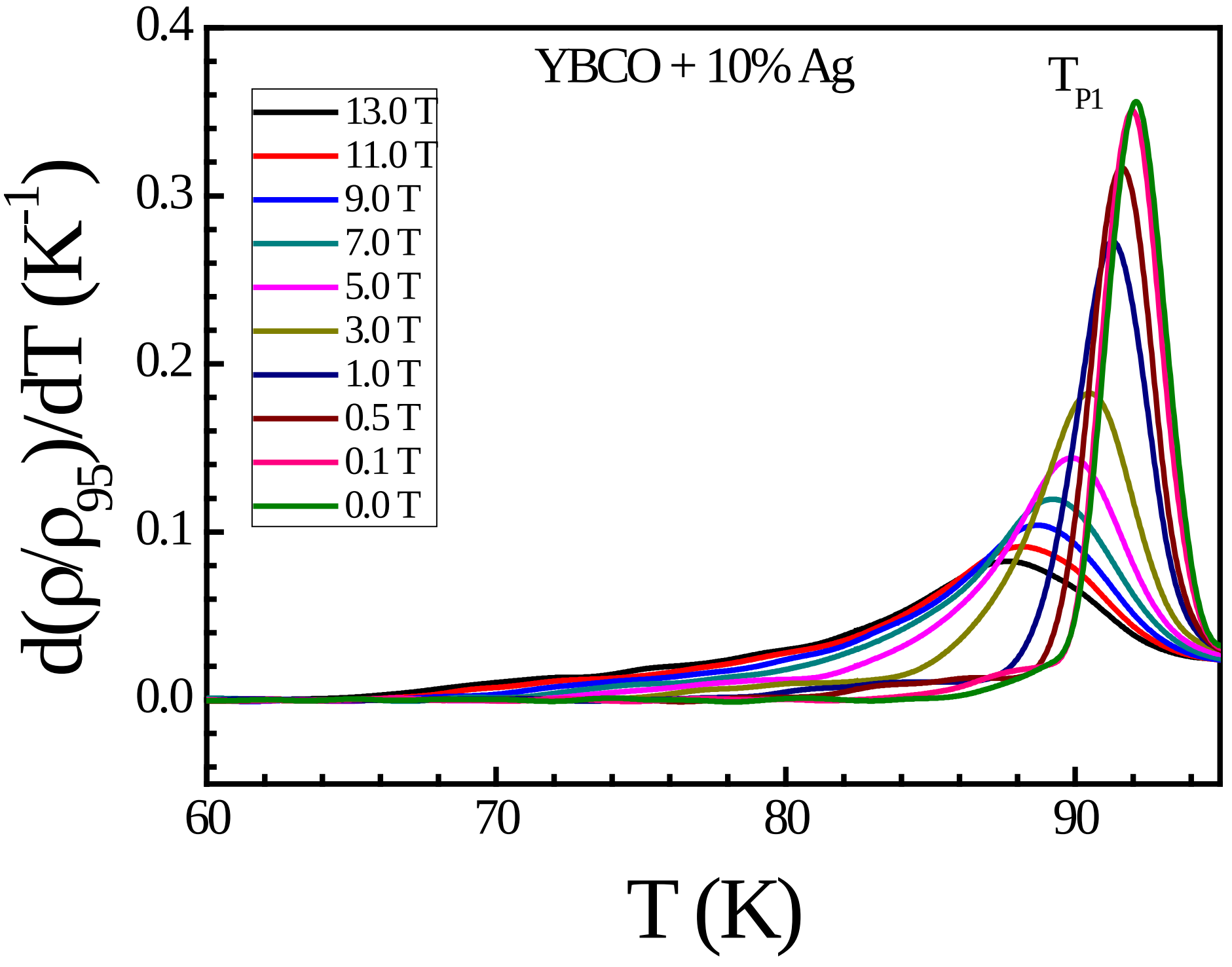

Figure 3 (c)

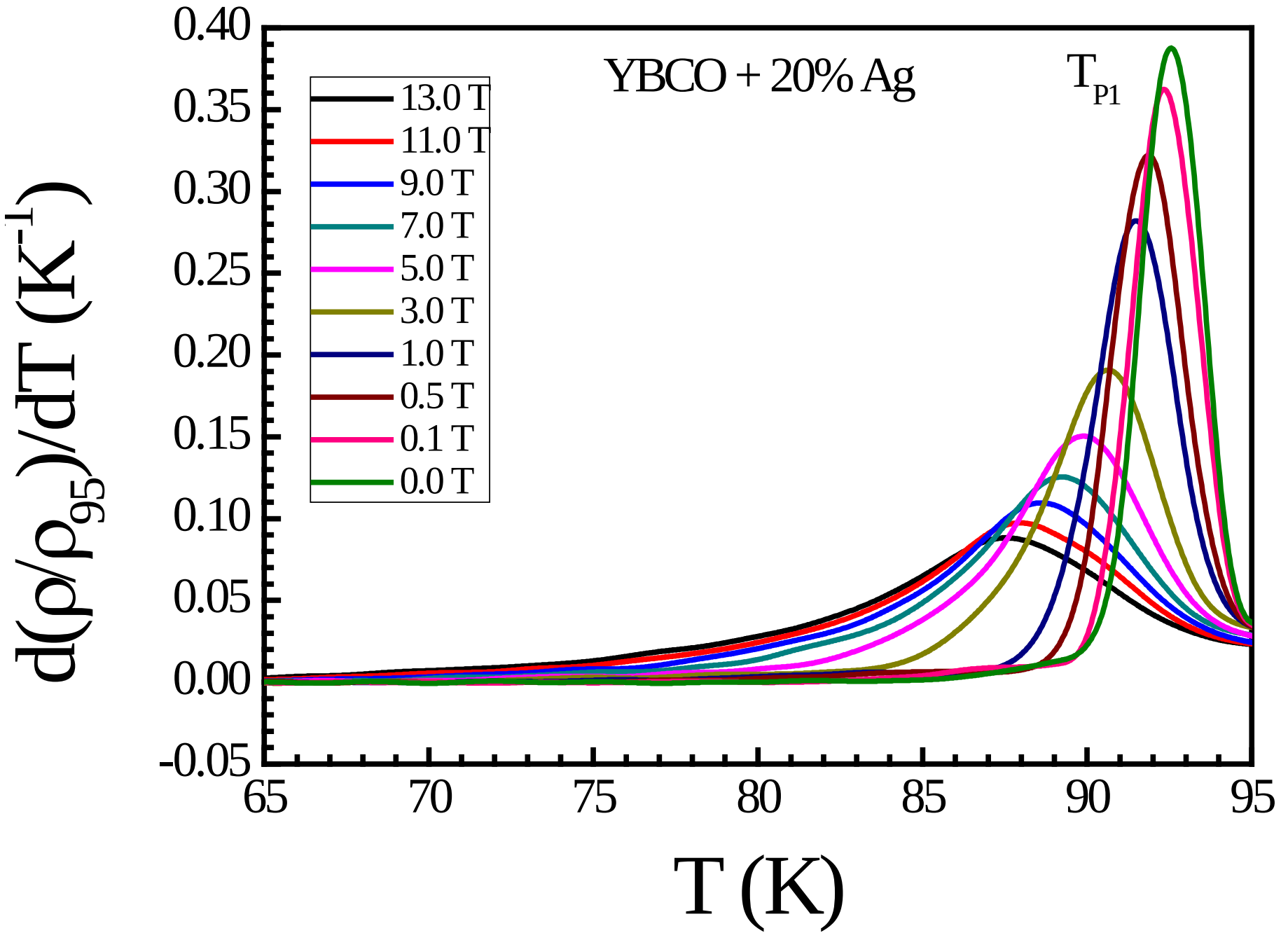

Figure 4

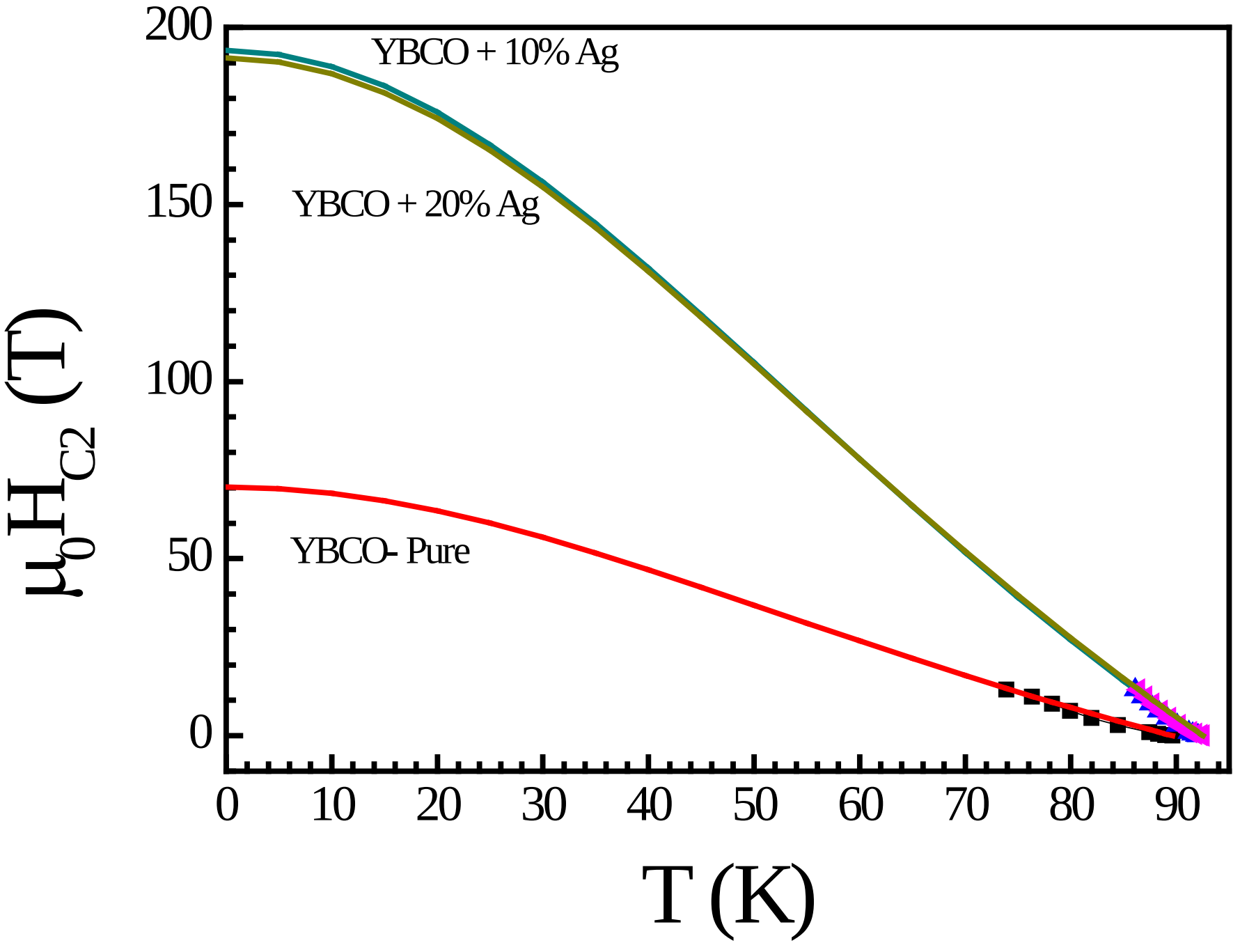

Figure 5 (a)

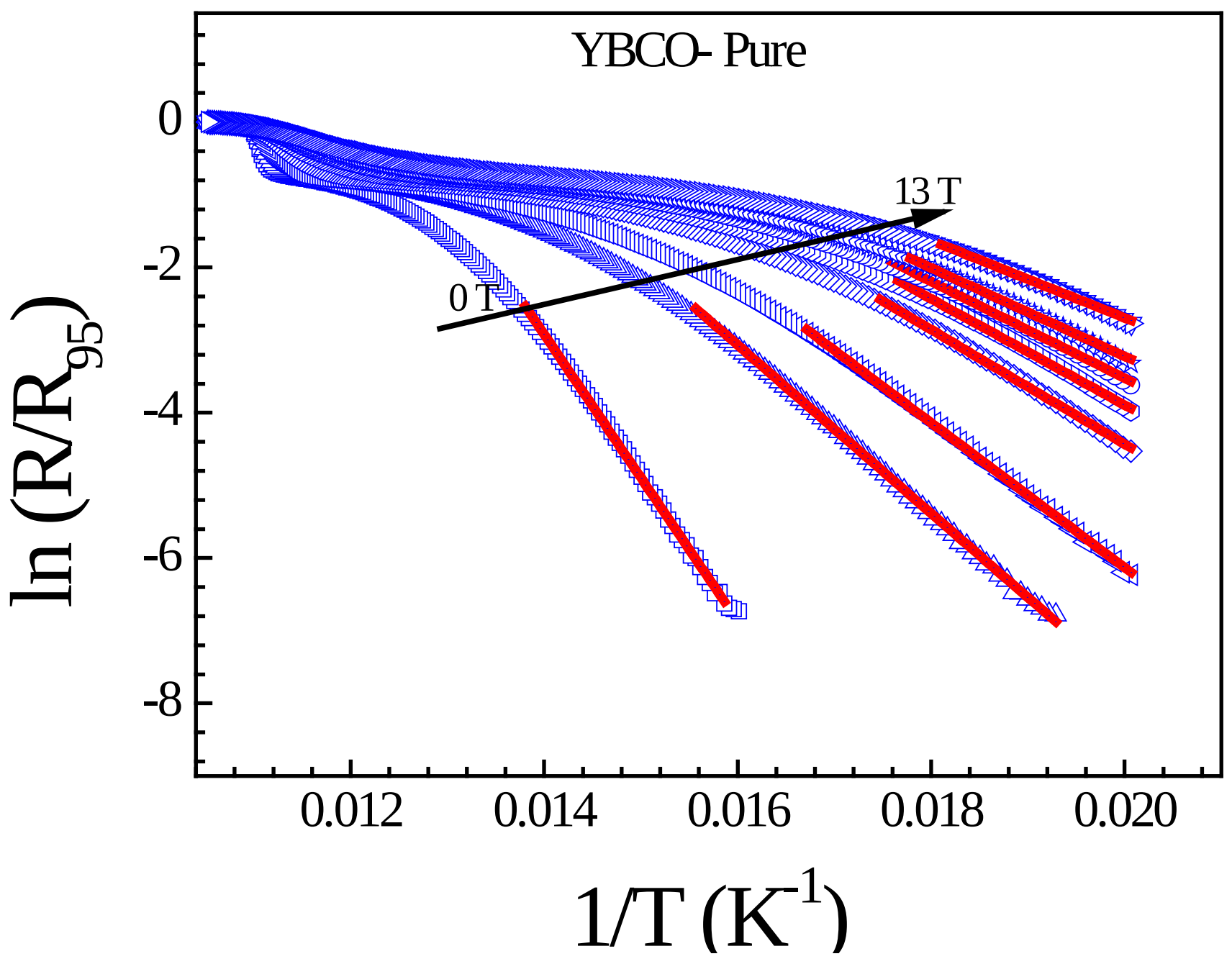

Figure 5 (b)

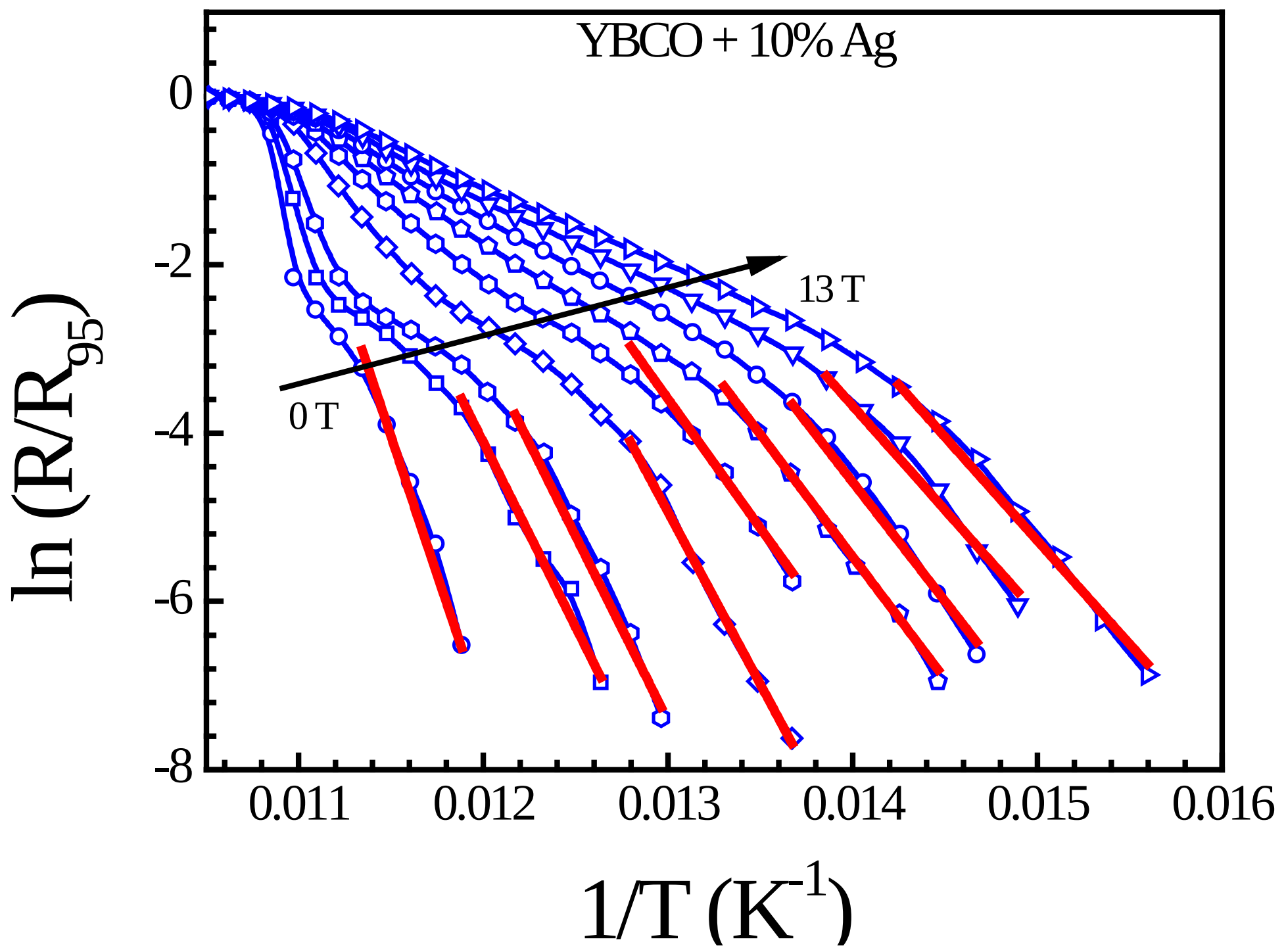

Figure 5 (c)

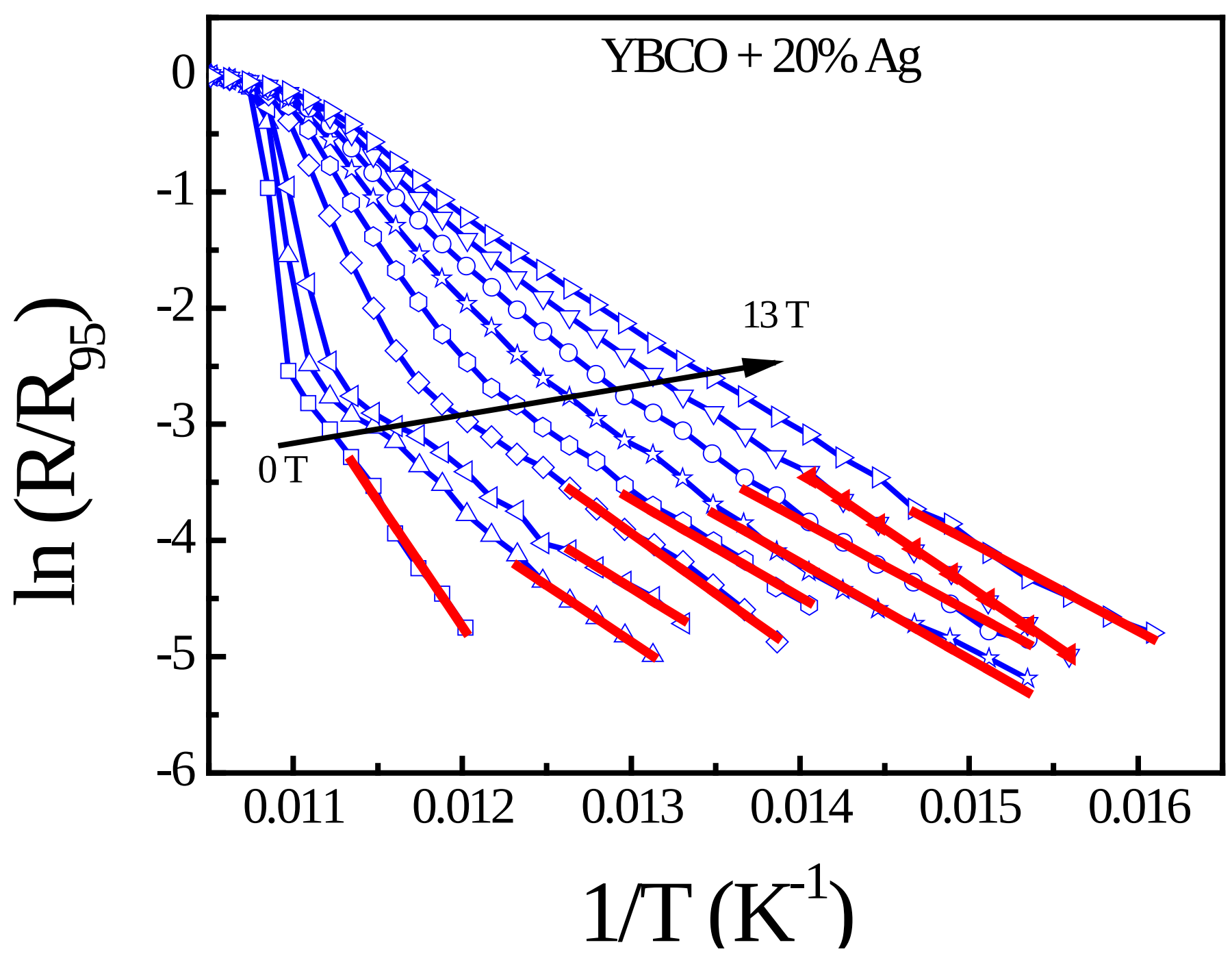

Figure 6 (a)

Figure 6 (b)

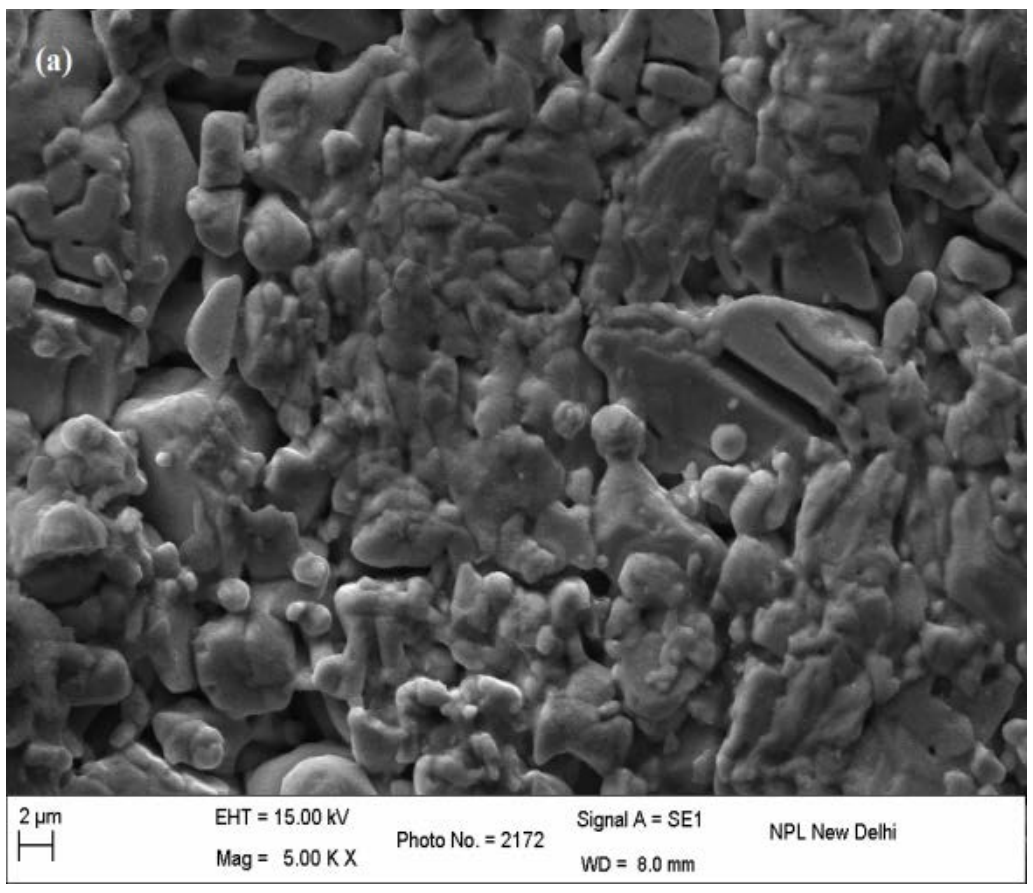

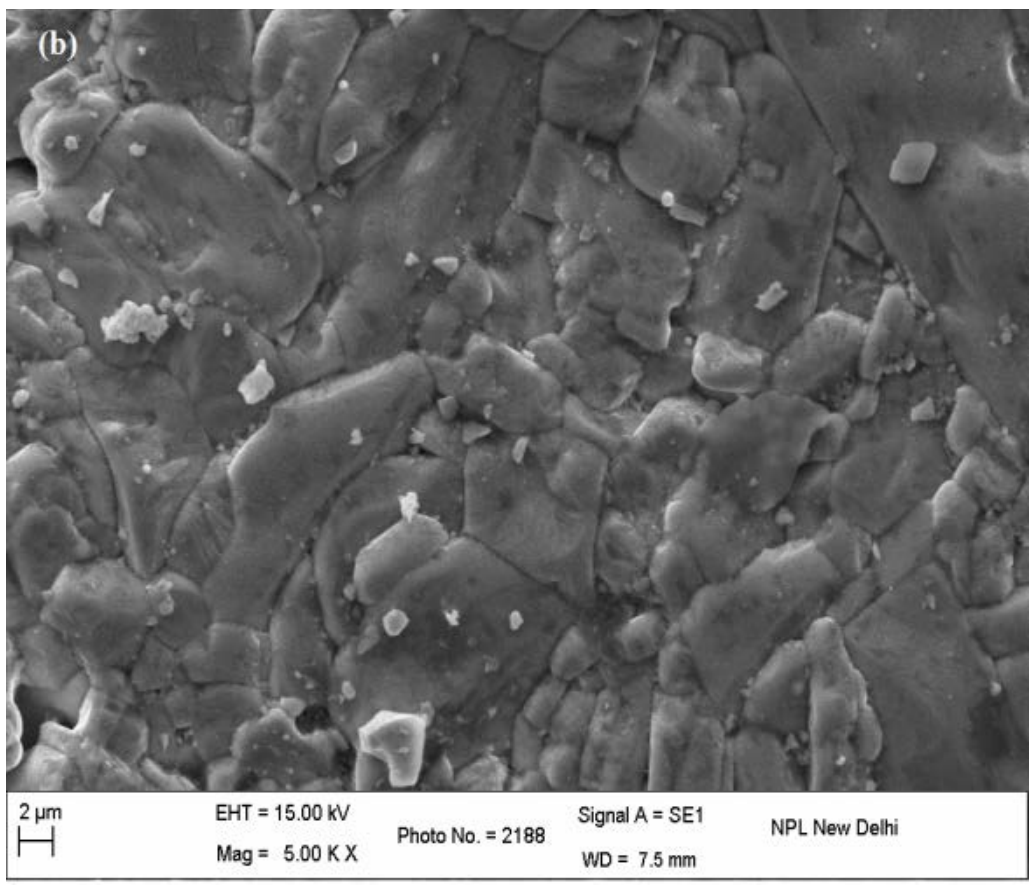